\documentclass{mem}
\usepackage{natbib}\usepackage{txfonts}\usepackage{balance}
\usepackage{graphicx}
\usepackage[a4paper,breaklinks,dvipdfm]{hyperref}
\usepackage{subfigure}

\begin{document}

\title{
Metal-free galaxy candidates discovered in CLASH
}

\author{
C-E. \,Rydberg\inst{1} 
          }

  \offprints{C-E. Rydberg}

\institute{
Department of Astronomy, Oscar Klein Center, Stockholm University, Albanova, Stockholm SE-106 91, Sweden
\email{claes-erik.rydberg@astro.su.se}
}

\authorrunning{Rydberg }

\titlerunning{Metal-free galaxy candidates}

\abstract{
The first metals in the universe are expected to form in population III stars -- primordial stars consisting entirely of hydrogen and helium. However, these stars have so far remained elusive. Simulations indicate that galaxies consisting exclusively, or almost exclusively, of population III stars may form at $z>6$, and such galaxies may provide one of the best probes of the properties of the population III star formation mode. By fitting Yggdrasil model spectra to multiband photometry data, we have identified four population III galaxy candidates in the Cluster Lensing and Supernova survey with Hubble (CLASH). We rule out alternative mundane galaxies and low redshift interlopers through similar fits to catalogs of spectra from more mundane objects. If confirmed through spectroscopy, this would constitute the first detection of the ``missing link'' between the early pristine universe and the metal-enriched universe.
}
\maketitle{}

\section{Introduction}

Simulations \citep[e.g.][]{2004ARA&A..42...79B} of the very first population III stars indicate that they formed in pristine mini-halos ($10^5-10^6$~M$_{\odot}$) at $z<30$. The stars are predicted to form as binary stars or small clusters \citep[e.g][]{2010MNRAS.403...45S, 2011ApJ...731L..38P, 2011ApJ...727..110C, 2011ApJ...737...75G, 2012MNRAS.422..290S, 2013RPPh...76k2901B, 2013arXiv1308.4456H} with mass of a few times 10~M$_{\odot}$. However, this first generation of population III stars are not likely to be detected even with the James Webb Space Telescope \citep{2013MNRAS.429.3658R}.

Chemically pristine dark matter halos may survive until much lower redshifts \citep{2010ApJ...716L.190S, 2011Sci...334.1245F} leading to population III galaxy formation at $z<15$. Since there are no clear-cut detections of population III galaxies yet (but see e.g. \citet{2011MNRAS.411.2336I, 2011MNRAS.418L.104Z, 2012ApJ...761...85K} for tantalizing candidates), their properties and possible observational signatures are highly uncertain. A lack of metal-lines are predicted \citep{2003A&A...397..527S, 2010A&A...523A..64R, 2011MNRAS.415.2920I, 2011ApJ...740...13Z}, especially the strong [OIII] (5007 $\AA$) line. Models also predict a strong HeII (1640~$\AA$) line and a steep UV slope \citep[e.g.][]{2011MNRAS.411.2336I, 2011ApJ...740...13Z}. The intrinsic Ly$\alpha$ line is also predicted to be very strong \citep[e.g.][]{2011MNRAS.418L.104Z}. The Ly$\alpha$ line is resonant and prone to scattering in the IGM and absorption of dust in the ISM \citep{2010Natur.464..562H}. However, it is plausible to assume that population III galaxies contain no dust since they lack metals to form dust. If a sufficient fraction Ly$\alpha$-photons also escape the IGM scattering (if, for example, a sufficiently large volume around the galaxy is ionized so the photons become sufficiently redshifted before encountering neutral hydrogen in the IGM), a strong Ly$\alpha$-line could constitute a signature of population III galaxies \citep{2011MNRAS.418L.104Z}. At $z>6$, the Gunn-Peterson trough \citep{1965ApJ...142.1633G} absorbs radiation with wavelength below Ly$\alpha$. This follows from Ly$\alpha$-scatter in the neutral IGM. Using this effect, the redshift of objects can be determined by examining in what filters (and thereby at what wavelength) an object is detected. This is the so-called Lyman-break technique.

Simulations indicate population III galaxies to be low-mass objects \citep{2010ApJ...716L.190S} and thereby faint. However, the magnification associated with gravitational lensing improves the prospects of detection significantly \citep{2012MNRAS.427.2212Z}. The Cluster Lensing and Supernova survey with Hubble (CLASH) observes gravitational lenses in the form of galaxy clusters and is therefore ideal for this purpose. This survey maps their dark-matter distribution but also tries to find gravitationally lensed high-redshift galaxies \citep{2012ApJS..199...25P}. We have fitted population III galaxy models as well as comparison empirical/synthetic models to CLASH observations to discover population III galaxies. Here we present the discovery of four candidates in different clusters. The focus on the proceedings will be on the candidate found in RXJ~1347.

\section{Method}
\label{sec:method}

The CLASH data consists of broadband photometry for 25 galaxy clusters in 16-17 filters, covering a wavelength range of 2,000~--~17,000 Å. The results are supplied as FITS images as well as catalogs of objects identified with Sextractor \citep{1996A&AS..117..393B}. We have used 21 of these official catalogs, containing a total of 44,784 objects.

To model the population III galaxies we use Yggdrasil \citep{2011ApJ...740...13Z}, a spectral synthesis model  which combines stellar population spectral energy distributions (SEDs) with the SED of the surrounding nebula to construct the total SED of the first galaxies. It uses several parameters (age, time of starburst forming the galaxy etc). We also use the combined escape fraction of Ly$\alpha$ through both the ISM and IGM ($f_\mathrm{Ly\alpha}$) as a parameter.

We use two model grids containing mundane objects as comparison models. Gissel \citep{2003MNRAS.344.1000B} is a commonly used grid of synthetic models. The second is the CWW, Kinney grid \citep{1999MNRAS.310..540A}. It is based on empirical spectra, but extrapolated into the infrared by models.

To fit the observations to the models we first use the Le Phare \citep{2006A&A...457..841I} code. The code is fast and provides us with a simple $\chi^2$-fit of the observations to the models. The $\chi^2$ is used to provide a rough scan of the observations to find a smaller set of potential candidates. We then use cross-validation \citep{Singh1981} on this smaller set. Cross-validation is a maximum-likelihood method producing a likelihood value for each fit. To perform cross-validation on an observation one filter is removed at a time. A $\chi^2$-fit is then calculated on the values in the remaining filters. The fit of the model then predicts a value for the missing filter. By comparing this value to the observed value in the filter a prediction error can be estimated. By repeating this for all filters, using a normal-distribution, a maximum likelihood value can be constructed for the observation. This cross-validation procedure together with closer examination of the FITS images is then used to filter the set of potential candidates.

\section{Results}

We used the procedure described in Section \ref{sec:method} to filter the 44,784 objects from the official catalogs. As a result of this we arrived at four candidates, one each in the clusters RXJ~1347, Abell~2261, MACS~1931, and MACS~0647. Table \ref{tab:candidateproperties} displays data for the four candidates. As can be seen in the table they have redshift estimates between 6.3 and 8.8. The lower limits of magnification is between 1.3 and 5.7 while the estimate for the Abell~2261 candidate is 8.1. The lower estimates are found by using estimates from \citet{2013arXiv1308.1692B} on galaxies that are further from the critical line than the candidates, thus they are expected to have a lower magnification. The candidate in Abell~2261 is actually one of the objects published in \citet{2013arXiv1308.1692B} providing us with an estimate. The estimates of the Ly$\alpha$ escape fractions vary between 40 to 80 \% but are very uncertain.

\begin{table}
\caption{Data for the four population III galaxy candidates. The candidates have extremely high redshift estimates ranging from 6.3 to 8.8. The magnification estimates are extracted from \citet{2013arXiv1308.1692B}. The Ly$\alpha$ estimates are the total escape fraction through both the ISM and the IGM. They are however highly uncertain.}
\label{tab:candidateproperties}
\begin{center}
\begin{tabular}{lccc}
\hline
\\
Cluster & $z$ & $\mu$ & $f_\mathrm{Ly\alpha}$ \\
\hline
\\
RXJ~1347  &$ 8.0 $ & $>5.7$& $50 \% $ \\
Abell~2261  &$ 6.3 $ & $\approx 8.1 $& $40 \% $ \\
MACS~1931  &$ 8.8 $ & $>1.3 $& $80 \% $ \\
MACS~0647  &$ 8.8 $ & $>4.7 $& $40 \% $
\\
\hline
\end{tabular}
\end{center}
\end{table}

\begin{figure}[t!]
\resizebox{\hsize}{!}{\includegraphics[clip=true]{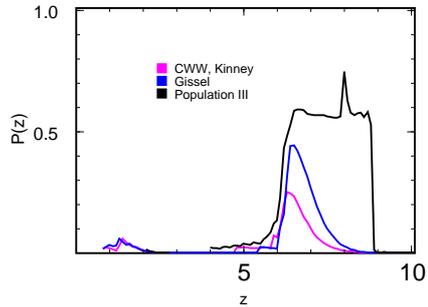}}
\caption{\footnotesize
The cross validation fitting result of model grids to observational data for the population III galaxy candidate in RXJ~1347 as a function of $z$. The y-axis indicate the accuracy of the cross-validation fit. A value of one would indicate an exact prediction of every removed observation. The population III models consists of the Yggdrasil galaxy grid. Gissel and CWW, Kinney are comparison grids of more mundane galaxies. For each redshift the best fitting model/parameter values in each grid are found. The population III fitting result dominates even though at least Gissel have good fits around $z=6.4$.
}
\label{fig:rxj1347cv}
\end{figure}

All of the four candidates have cross-validation fits that are significantly better than the comparison models for mundane galaxies. Concentrating on the candidate in RXJ~1347 the fit as a function of redshift is displayed in Figure \ref{fig:rxj1347cv}. There is a quite broad range with good fit for $z$ between $6.2$ and $8.8$ with a sharp peak at $z=8.0$. The comparison models generally have much worse fits even though there are a peak around $z=6.4$ where at least a Gissel model fits rather well.

\begin{figure*}[htp]
  \begin{center}
	  \subfigure[F775W]{\includegraphics[scale=0.075]{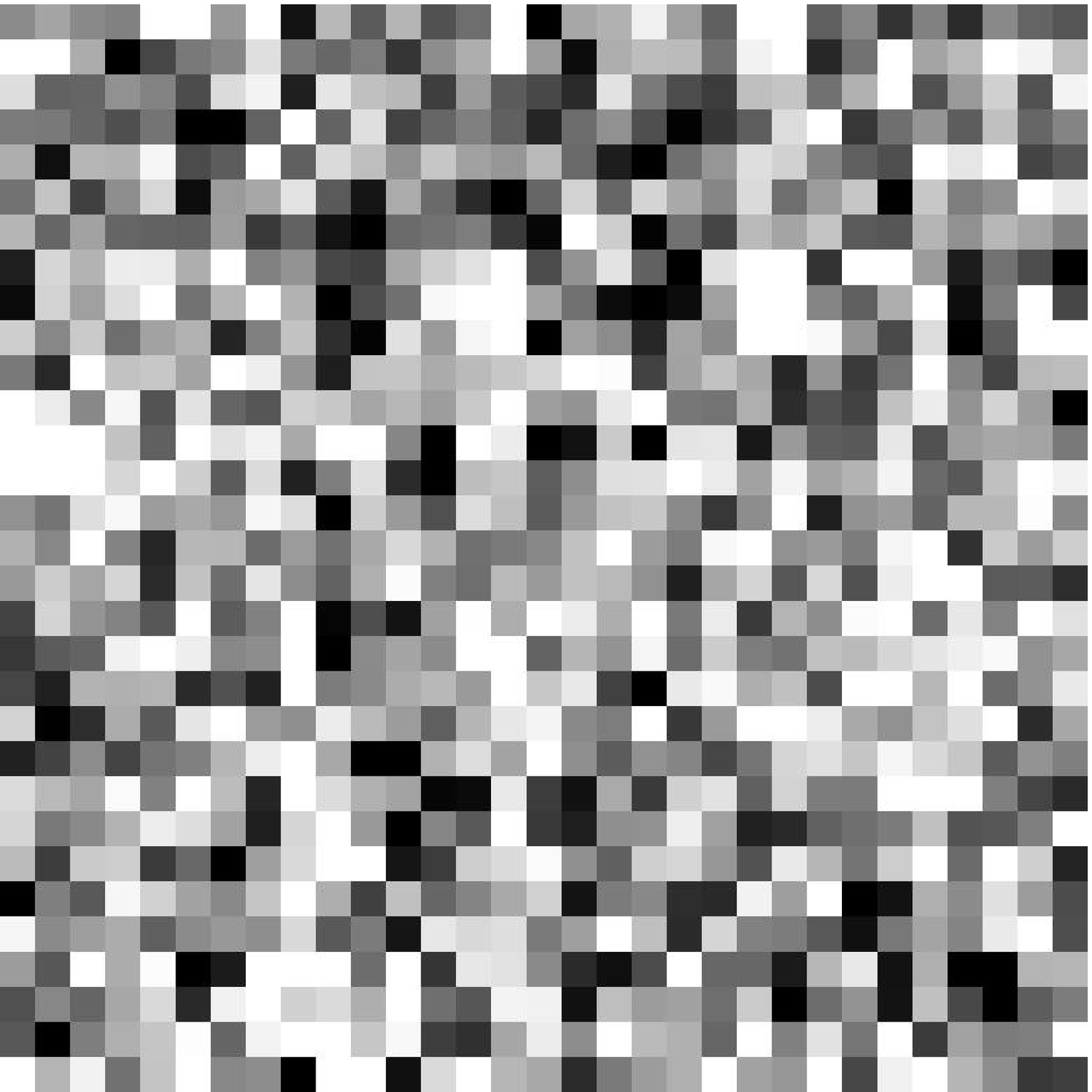}}
		\subfigure[F814W]{\includegraphics[scale=0.075]{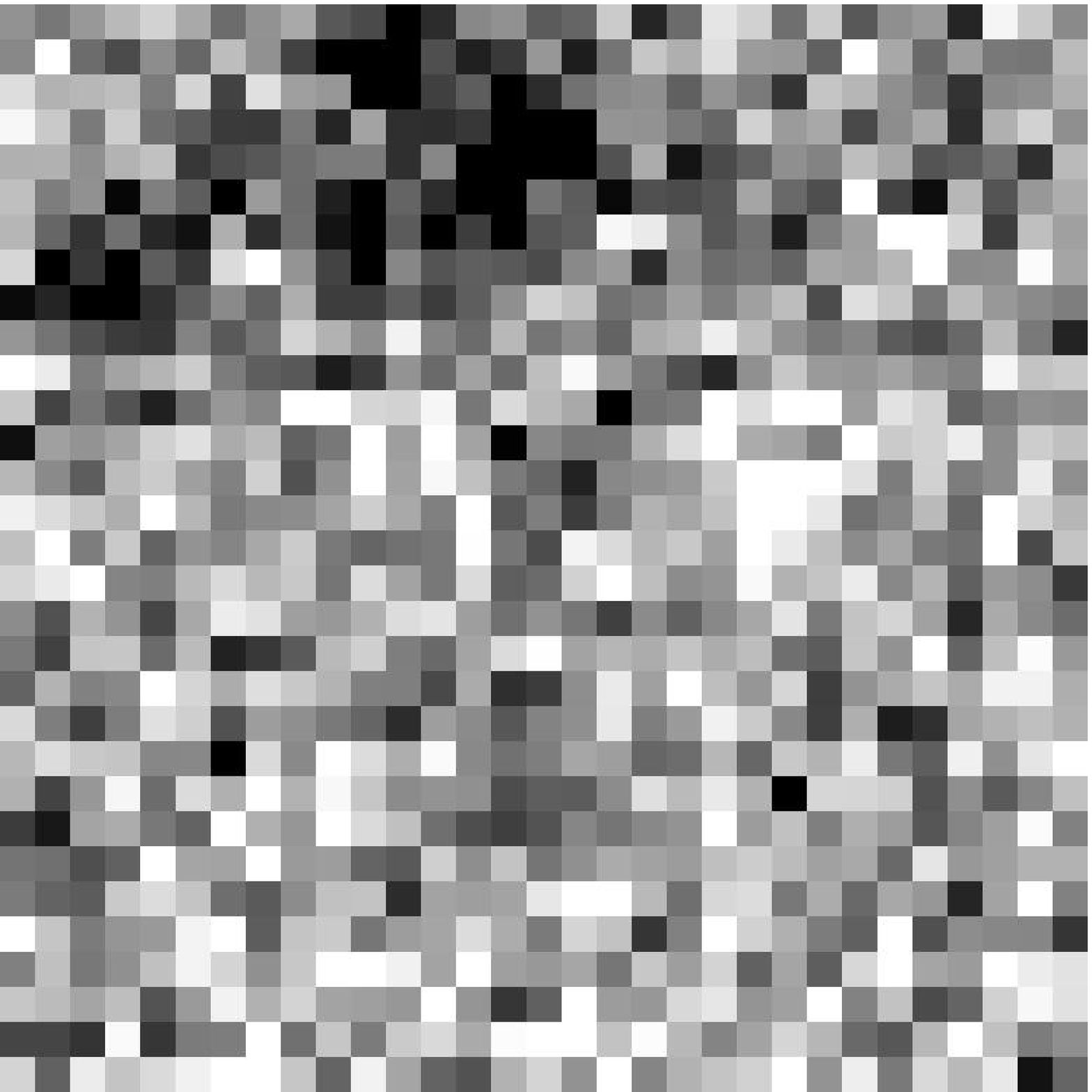}}
		\subfigure[F850LP]{\includegraphics[scale=0.075]{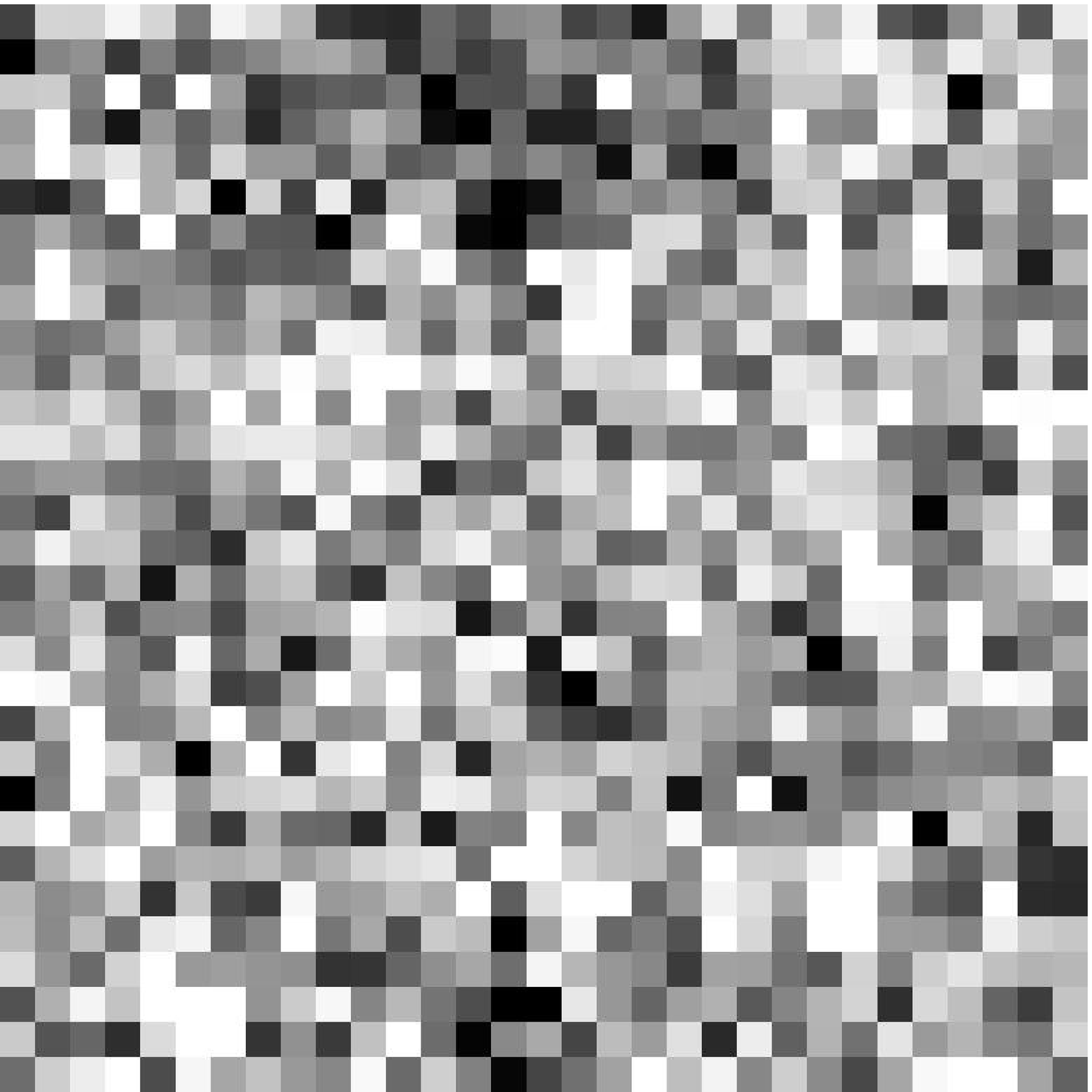}}
		\subfigure[F105W]{\includegraphics[scale=0.075]{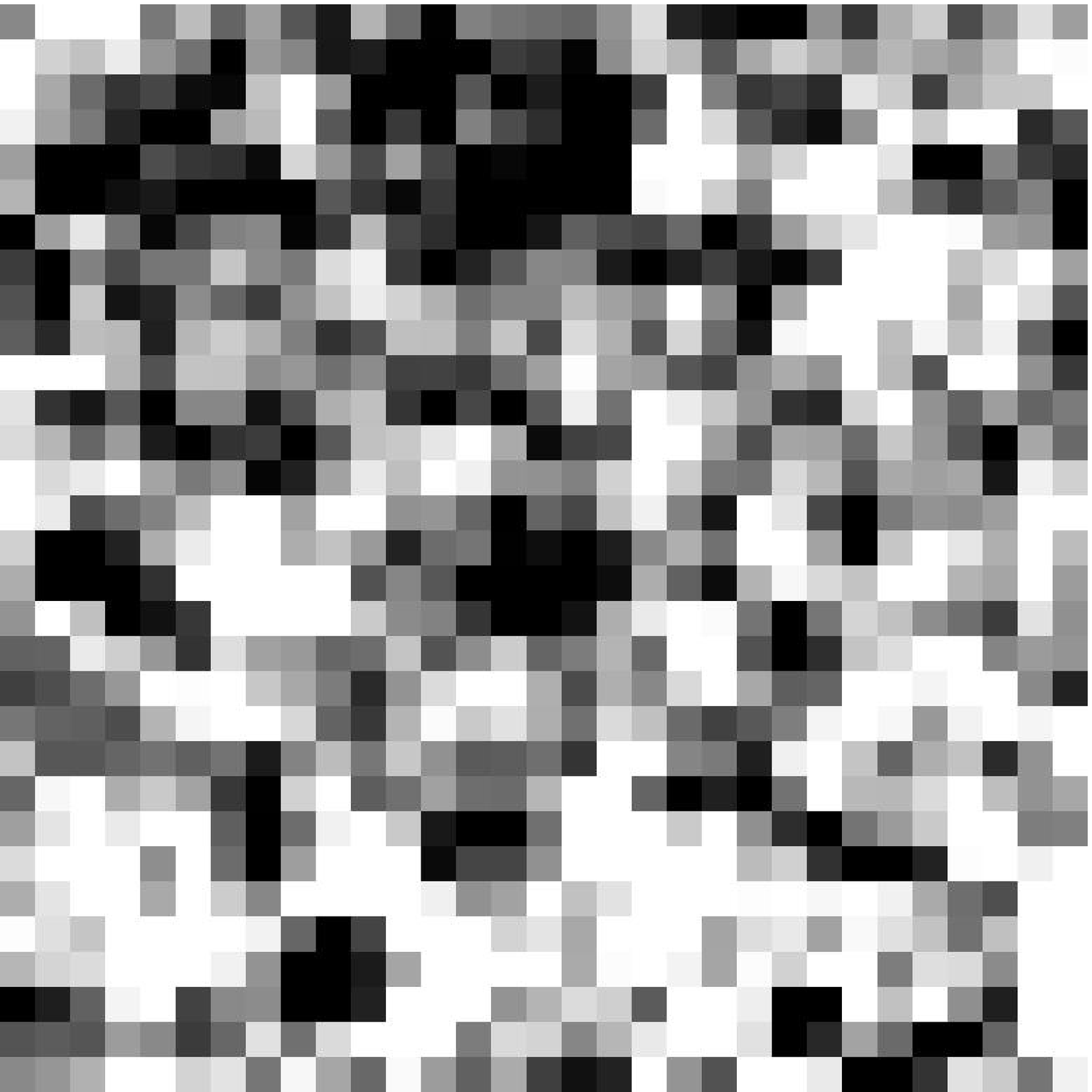}}
		\subfigure[F110W]{\includegraphics[scale=0.075]{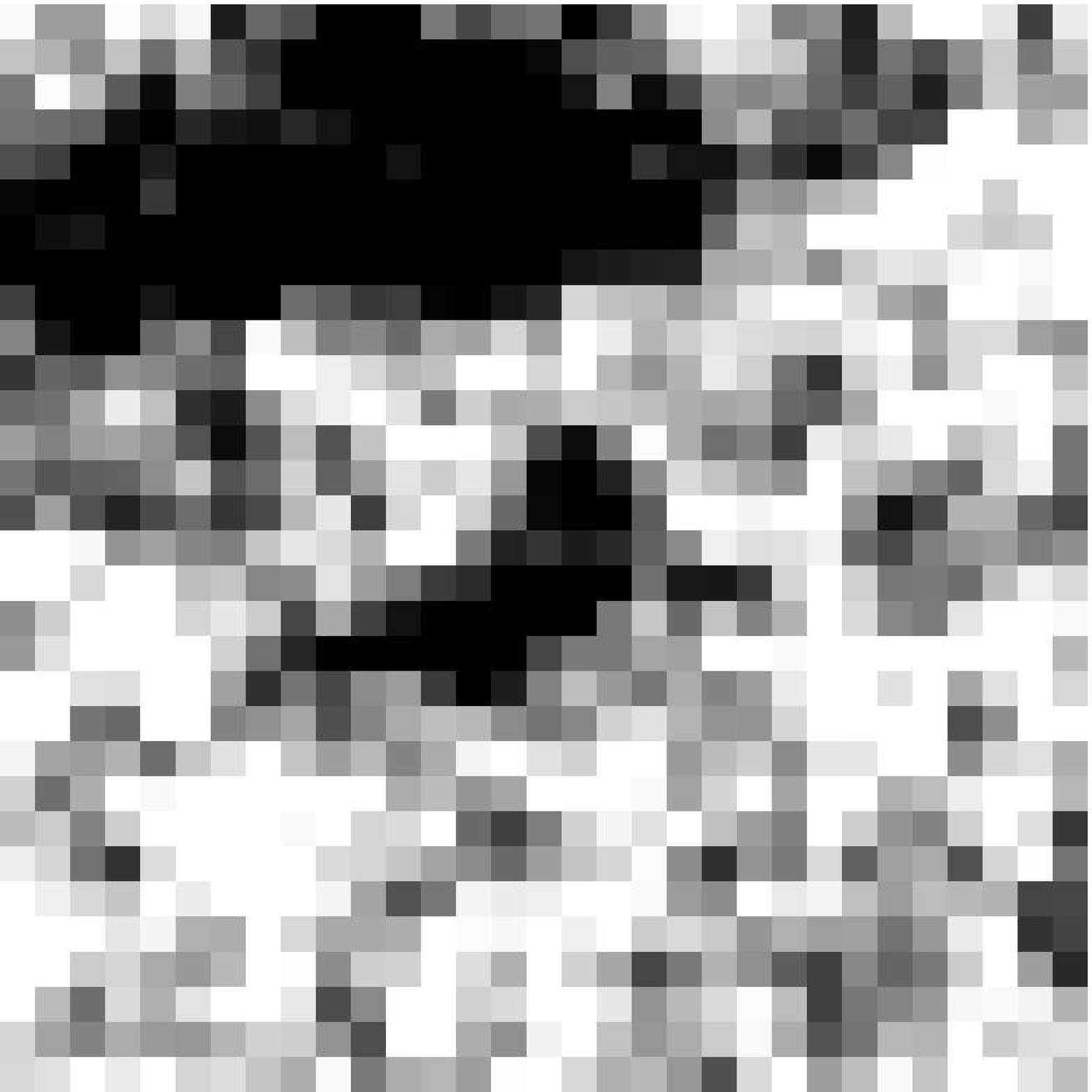}}
		\subfigure[F125W]{\includegraphics[scale=0.075]{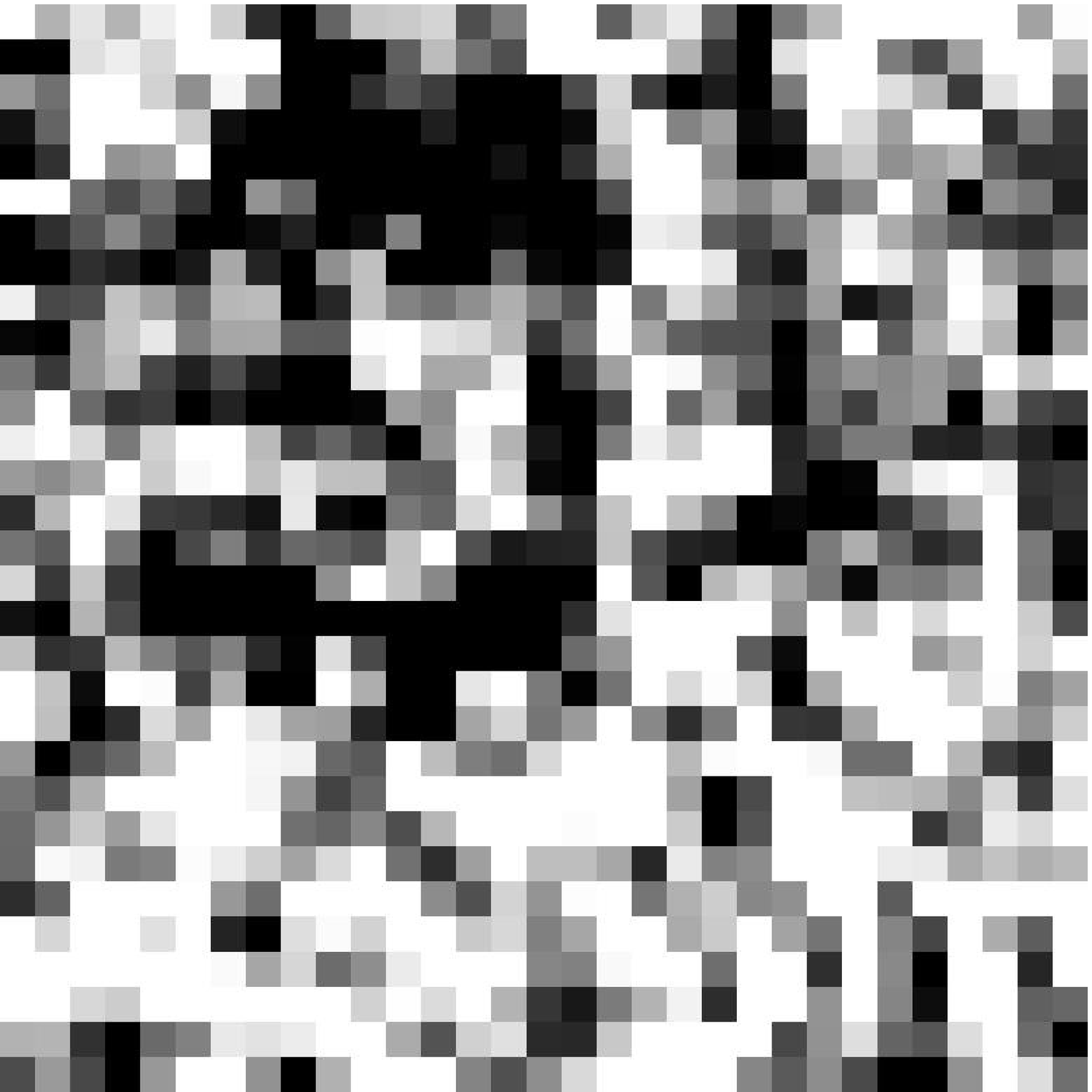}}
		\subfigure[F140W]{\includegraphics[scale=0.075]{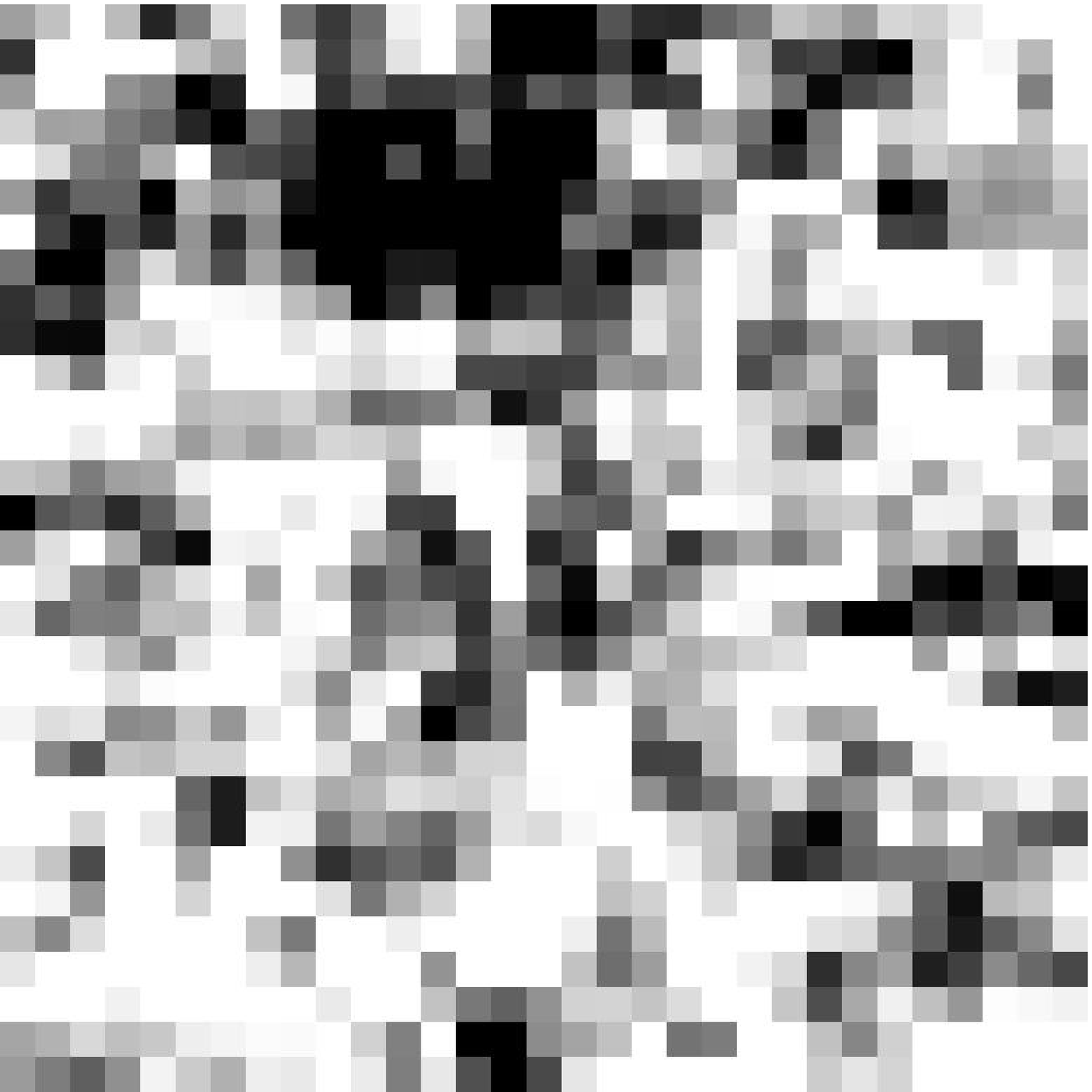}}
		\subfigure[F160W]{\includegraphics[scale=0.075]{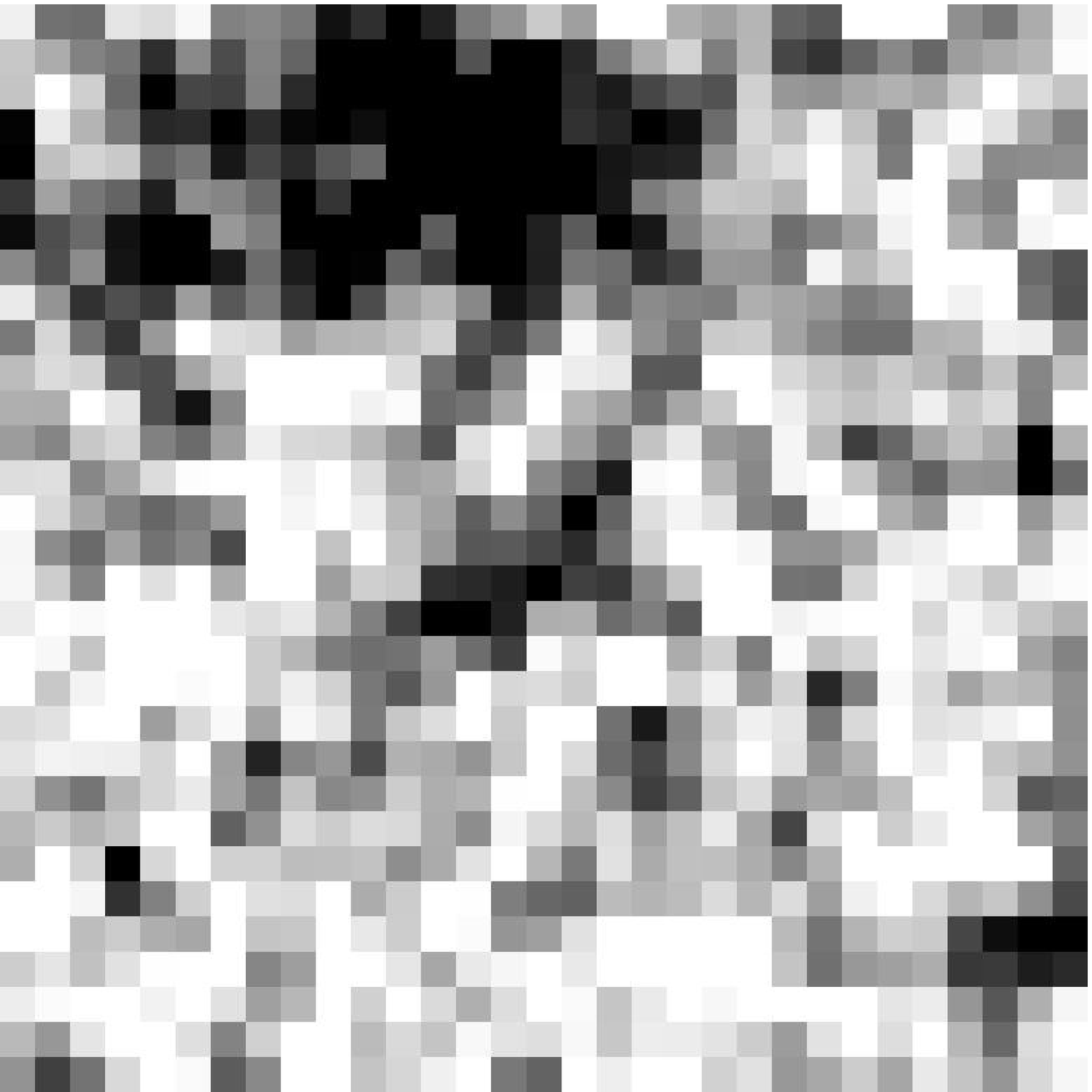}}
  \end{center}
  \caption{\footnotesize
$2.0'' \times 2.0''$ thumbnail images of the population III galaxy candidate (centered in the image) in RXJ~1347. The images are from the eight CLASH filters covering the longest wavelengths, ranging between $6,500 \AA$ and $17,000 \AA$. The Lyman-break is visible between F850LP and F105W since the object is observed in F105W but not in F850LP. The slight distortion from upper right to lower left that can be seen in the object (at least in some of the longer wavelength filters) could imply gravitational lensing with high magnification.}
  \label{fig:rxj1347thumbnails}
\end{figure*}

The Figure \ref{fig:rxj1347thumbnails} shows $2.0'' \times 2.0''$ thumbnail images of the candidate in RXJ~1347. The eight filters in CLASH covering the longest wavelengths are used, ranging from 6,500~$\AA$~(F775W) to 17,000~$\AA$~(F160W). The Lyman-break is clearly visible between F850LP and F105W since the object is observed in F105W but not in F850LP. In the filters F110W (which contains the most significant detection with S/N=8.3) to F160W the galaxy appears slightly distorted, seemingly stretched from upper right to lower left. This could imply gravitational lensing with high magnification.

\section{Conclusion}

By fitting the observations from CLASH to models of population III galaxies and mundane comparison galaxies we have extracted four population III galaxy candidates. Three of the candidates are the first $z>7.0$ population III galaxy candidates discovered. The candidate in RXJ~1347 also show signs of being gravitationally lensed by being slightly distorted. However, the parameter dependence of the population III grid needs to be explored more thoroughly. The escape fraction of Ly$\alpha$ photons, where a very uncertain estimate is provided in this proceeding, needs more examination. More precise magnification estimates should also be derived. This together with a careful examination of the cross-validation results should provide us with estimates of the mass of the candidates. And finally, the same procedure should be used on the four clusters in CLASH we have not examined yet.
 
\begin{acknowledgements}

CER acknowledge funding from the Swedish National Space Board and the Royal Swedish Academy of Sciences.

\end{acknowledgements}

\bibliographystyle{aa}
\bibliography{References}

\end{document}